\documentclass[runningheads]{llncs}
\usepackage{bm}
\usepackage{algorithm}
\usepackage{algorithmic}
\usepackage{comment}
\usepackage{microtype}
\usepackage{enumitem}
\usepackage{booktabs}

\usepackage{graphicx}
\usepackage{amsfonts}
\usepackage{amsmath}
\usepackage{amssymb}
\usepackage{subfigure}
\usepackage{graphbox}
\usepackage{multirow}
\usepackage[misc]{ifsym}

\begin{document}
	\title{Customized Conversational Recommender Systems}
	
	\author{Shuokai Li\inst{1,2,}\thanks{The authors are at the Key Lab of Intelligent Information Processing of Chinese Academy of Sciences. \Letter\  indicates corresponding authors.} \and Yongchun Zhu\inst{1,2,\star} \and Ruobing Xie\inst{3} \and Zhenwei Tang\inst{4} \and Zhao Zhang\inst{1} \and Fuzhen Zhuang\inst{5,6} \Letter \and Qing He\inst{1,2,\star} \Letter \and Hui Xiong\inst{7}}
	
	\institute{\textls[-16]{Institute of Computing Technology, Chinese Academy of Sciences, Beijing 100190, China} \and University of Chinese Academy of Sciences, Beijing 100049, China \and WeChat Search Application Department, Tencent, China \and King Abdullah University of Science and Technology, Thuwal, Saudi Arabia \and Institute of Artificial Intelligence, Beihang University, Beijing 100191, China \and SKLSDE, School of Computer Science, Beihang University, Beijing 100191, China \and Artificial Intelligence Thrust, The Hong Kong University of Science and Technology, Guangzhou, China \\ \email{\{lishuokai18z,zhuyongchun18s,zhangzhao2021,heqing\}@ict.ac.cn \\ ruobingxie@tencent.com zhenwei.tang@kaust.edu.sa \\ zhuangfuzhen@buaa.edu.cn xionghui@ust.hk}}
	
	\maketitle 
	
	\begin{abstract}
		\textls[-0]{Conversational recommender systems (CRS) aim to capture user's current intentions and provide recommendations through real-time multi-turn conversational interactions.
		As a human-machine interactive system,
		it is essential for CRS to improve the user experience.
		However, most CRS methods neglect the importance of user experience. 
		In this paper, we propose two key points for CRS to improve the user experience: 
		(1) \textit{Speaking like a human}, human can speak with different styles according to the current dialogue context. 
		(2) \textit{Identifying fine-grained intentions}, even for the same utterance, different users have diverse fine-grained intentions, which are related to users' inherent preference.
		Based on the observations, we propose a novel CRS model, coined \textbf{C}ustomized \textbf{C}onversational \textbf{R}ecommender \textbf{S}ystem (CCRS), which customizes CRS model for users from three perspectives. 
		For human-like dialogue services, we propose multi-style dialogue response generator which selects context-aware speaking style for utterance generation.
		To provide personalized recommendations, we extract user's current fine-grained intentions from dialogue context with the guidance of user's inherent preferences. 
		Finally, to customize the model parameters for each user, we train the model from the meta-learning perspective. 
		Extensive experiments and a series of analyses have shown the superiority of our CCRS on both the recommendation and dialogue services. }
	\end{abstract}
	
	\section{Introduction}

	Recently, conversational recommender systems (CRS)~\cite{zhang2018towards,sun2018conversational,li2018towards,chen2019towards,zhou2020improving,liu2020towards,li2022user}, which capture the user current preference and recommend high-quality items through real-time dialog interactions, have become an emerging research topic.
	They~\cite{li2018towards,chen2019towards,zhou2020improving} mainly require a dialogue system and a recommender system. The dialogue system elicits the user intentions through conversational interaction and responses to the user with reasonable utterances. On the other hand, the recommender system provides high-quality recommendations by user's intentions and inherent preferences.
	CRS has not only a high research value but also a broad application prospect~\cite{gao2021advances}, such as ``Siri", ``Cortana" etc. 
	
	\begin{figure*}[htpb]
		\centering  
		\includegraphics[width=0.95\textwidth]{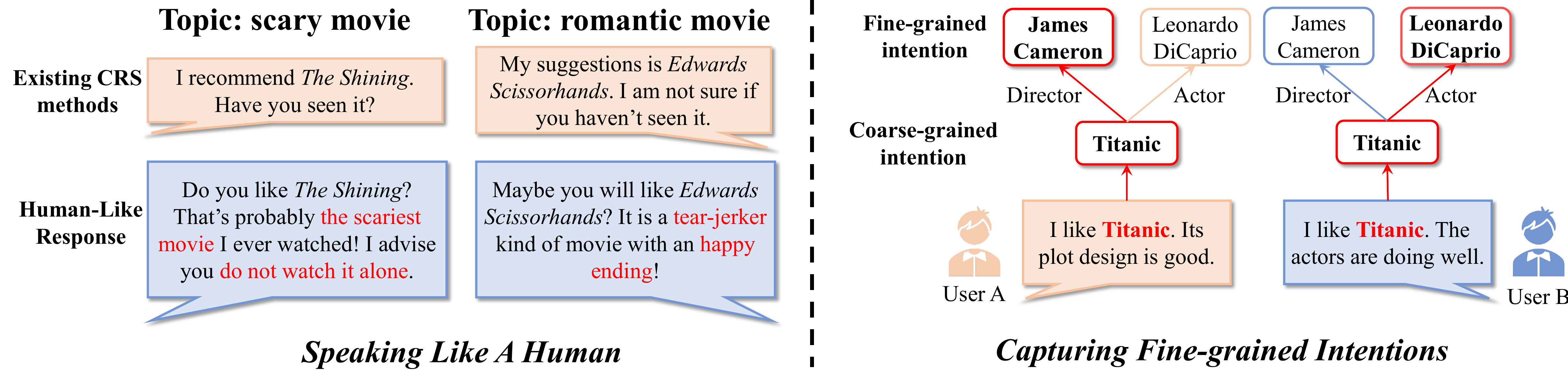}
		\caption{An motivating example of our CCRS. The left part shows the various human speaking styles on different topics. The right part presents different users have the same coarse-grained intentions but with various fine-grained intentions.
		}\label{framework}
		\vspace{-1mm}
	\end{figure*}
	
	As a kind of human-machine interactive system, improving user experience is of vital importance. 
	However, existing CRS methods neglect the importance of user experience. Some methods~\cite{zhang2018towards,sun2018conversational} not only require lots of labor to construct rules or templates but also make results rely on the pre-processing, which hurt user experience as the constrained interaction~\cite{gao2021advances}. Some other approaches~\cite{li2018towards,chen2019towards,zhou2020towards,zhou2020improving,lu2021revcore} generate inflexible and fixed-style responses which could make users uncomfortable. Besides, some methods~\cite{chen2019towards,zhou2020improving} only identify coarse-grained intentions and cannot provide customized recommendations. 
	In this paper, to improve the user experience in CRS, we propose two key points:
	\begin{itemize}[leftmargin=*]
		\item \textbf{\textit{Speaking like a human:}} \textls[-16]{Facing various dialogue scenes, people’s responses may be diverse largely in terms of speaking styles. 
		Figure~\ref{framework} (left) shows an example when the topics are about horror films and romantic movies.
		Obviously, a human-like dialogue system is expected to: (1) generate utterances that fit the current content semantics and topics, rather than using a fixed template; (2) generate vivid and attractive conversations, rather than short, dull and boring expressions.
		In this way, the user experience would be improved and user engagement would also increase, which helps identify the user intentions more accurately.}
		\item \textbf{\textit{Identifying fine-grained intentions:}} The same utterance from different users could reflect diverse fine-grained intentions. For example, in Figure~\ref{framework} (right), if users mentioned the movie ``\textit{Titanic}'' during the conversation, they would have the same coarse-grained intentions to find movies related to ``\textit{Titanic}''. Nevertheless, they may have different fine-grained intentions.
		This is because of the diversity of the inherent preferences: some users prefer movies of the actor ``\textit{Leonardo DiCaprio}'', while others prefer movies of the director ``\textit{James Cameron}''. Thus, modeling the fine-grained intentions which are related to the users' inherent preferences helps provide high-quality recommendations.
	\end{itemize}
	
	\textls[-28]{Along this line, we design a novel model \textbf{C}ustomized \textbf{C}onversational \textbf{R}ecommender} \textbf{S}ystem (CCRS), which customizes CRS model from three perspectives.
	Firstly, for the recommender service,
	given user mentioned entities, 
	the key idea is to highlight user fine-grained intentions with the guidance of user's inherent preferences~(i.e. the preferences on different relations of entity).
	Then, for the dialogue service, we generate customized utterances with the guidance of content semantics.
	In detail, multiple styles are modeled as style embeddings, and we aggregate multiple style embeddings into the customized speaking style embedding according to the dialogue context. 
	Finally, we further customize the (recommendation and generation) models for each user, with the advantage of meta-learning.
	As user fine-grained intentions and speaking styles are sparse in CRS, we choose Model-Agnostic Meta-Learning (MAML) algorithm~\cite{finn2017model}, which can rapidly learn customized model parameters. 
	
	To summarize, the contributions of this paper are as follows:
	\begin{itemize}[leftmargin=*]
		\item \textls[-12]{We propose a novel customized conversational recommender system CCRS, which consists of customized recommendation and dialog services. We further customize the model parameters for each user with the advantage of meta-learning.}
		\item \textls[-18]{We model user fine-grained intentions on entity relations, and propose multi-style generation to provide human-like dialogue service, which improves user experience.}
		\item \textls[-12]{Extensive quantitative and qualitative experiments demonstrate that the proposed approach can significantly outperform baseline methods.}
	\end{itemize}

	\section{Method}
\begin{figure*}[htpb]
		\centering  
		\includegraphics[width=0.98\textwidth]{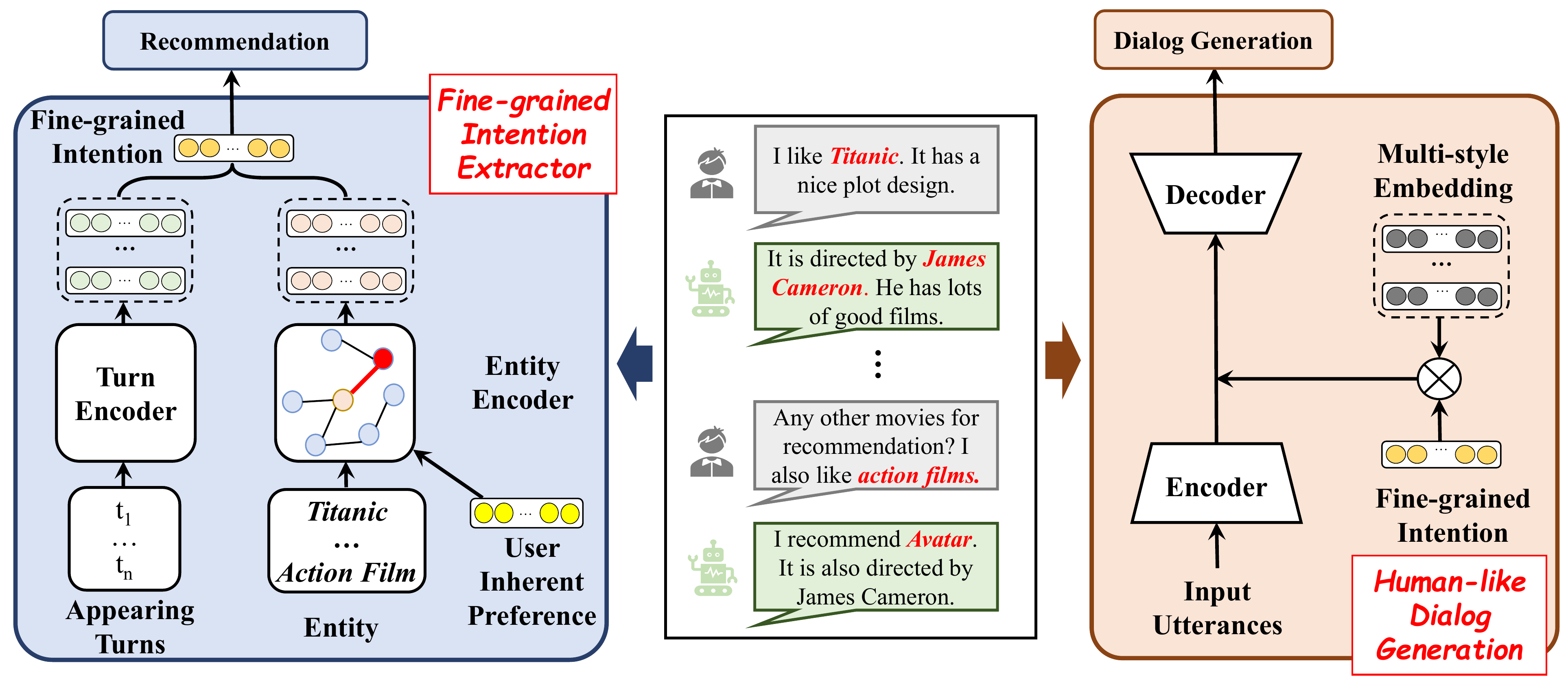}
		\caption{\textls[-0]{The overview of our CCRS, which consists of a customized recommender part and a personalized dialogue generation part. Moreover, we use MAML to learn customized model parameters for both recommendation and dialogue generation modules.
		}}\label{network}
		\vspace{-8mm}
	\end{figure*}

	In this section, we present our novel method to provide customized services for users, coined \textbf{C}ustomized \textbf{C}onversational \textbf{R}ecommender \textbf{S}ystem (CCRS). First, we extract the fine-grained intentions of users with the guidance of user inherent preferences.
	Then, we design the multi-style embeddings and generate personalized responses based on the extracted fine-grained user intentions.
	Finally, we adopt meta training to learn customized model parameters for each user rapidly. The overview illustration of the proposed model is presented in Figure~\ref{network}.
	
	\subsection{Preliminary and Formulations}
	\subsubsection{Recommender Module in CRS}
	Given a user $u\in\mathcal{U}$ with his identifier \textbf{$u_{id}$} and his mentioned entities $e\in\mathcal{E}$, a recommender system aims to retrieve a subset of items that meet the customized user needs. 
	To be noticed, the entities consist of item entities and non-item entities. For example, in a movie recommender system, the item entities denote the movies and the non-item entities can be the actor/actress, director, and genre. 
	To better model user mentioned entities, external knowledge graphs $\mathcal{G}$ are often incorporated.
	
	Specifically, the knowledge graph $\mathcal{G}$ consists of triples $(e, r, e')$ where the entities $e, e'\in\mathcal{E}$ and the entity relation $r\in\mathcal{R}$. $\mathcal{E}$ and $\mathcal{R}$ denote the entity set and entity relation set, respectively. Following~\cite{chen2019towards,zhou2020improving}, we utilize the knowledge graph from DBpedia~\cite{lehmann2015dbpedia} to learn the entity representations. As the original graph consists of redundant information, we collect all the entities appearing in the dialogue corpus and extract the subgraph following ~\cite{chen2019towards,zhou2020improving}.
	
	\subsubsection{Dialogue Module in CRS}
	The dialogue system is designed to generate proper utterance responses in natural languages. At the $T$-th turn of the conversation, the dialogue system receives the dialogue history $\mathcal{H}=\{s_t\}_{t=1}^{T-1}$ and user mentioned entities $\{e_1,e_2,...,e_n\}$. Here the entities are extracted from the dialogue utterances using entity linking. For simplify, the items are replaced by a special token ``$\textless$unk$\textgreater$". When generating ``$\textless$unk$\textgreater$", it means to provide recommendations for users. Then the items are provided by the recommendation part.

	\subsection{Fine-grained User Intentions Extraction}
	In this section, we extract the fine-grained user intentions by considering the entity relations and the appearing turns of the entities. These two factors are leveraged to learn the importance of user mentioned entities.
	
	
	
	\subsubsection{Entity Encoder}\label{attn}
	Given the user mentioned entities and corresponding knowledge graph, previous methods~\cite{chen2019towards,zhou2020improving} mainly leverage Relational Graph Convolutional Networks (R-GCNs)~\cite{schlichtkrull2018modeling} to incorporate the structural information and learn entity representations. 
	Though R-GCNs keep a distinct linear projection for each type of entity relation, the projections are fixed for different users, regardless of the fine-grained user intentions. 
	Indeed, the fixed projections could only model the coarse-grained preferences~(i.e. \textit{knowing what entities do users like}), and ignore the customized fine-grained user intentions on entity relations~(i.e. \textit{not knowing what types of entity relations do users like}).

	Motivated by~\cite{hu2020heterogeneous}, we proposed to capture the fine-grained user intentions by modeling personalized preference on entity relations. 
	For each user, with the guidance of the user inherent embedding, we learn the customized attention weights on various relations and aggregated the neighboring entities according to the customized attention weights.
	
	
	Given the triple $(e, r, e')$ in knowledge graph, where $e$ is the user mentioned entity, $e'\in N(e)$ is the neighbour of $e$ and $r$ is the relation between $e$ and $e'$, our goal is to learn the fine-grained user attentions on relation $r$. First, to learn the fine-grained entity information, we project the $l$-th layer's entity embeddings ($h^{l-1}(e)\in \mathbb{R}^{d\times 1}$) into multi-head representations:
	\begin{equation}
	T^{l-1}_i(e) = W^T_i h^{l-1}(e), \ \ \ \ \ \ \ \ \ \  S^{l-1}_i(e') = W^S_i h^{l-1}(e'),
	\end{equation}
	where $W^T_i\in \mathbb{R}^{\frac{d}{k}\times d}$ and $W^S_i\in \mathbb{R}^{\frac{d}{k}\times d}$ are trainable weights, and $h^0(e)={\rm Emb}_e(e)$ is initialized randomly. Next, as different users may have various inherent preferences, we keep a distinct relation-aware matrix $A_r^i$ and use $\gamma_{<r, u>}^i$ to represent the user intentions on distinct entity relation $r$. Concretely, $\gamma_{<r, u>}^i = {\rm Vec}(A_r^i)W_u \bm{U}_{id}$ is the similarity between $A_r^i$ and the inherent user embedding $\bm{U}_{id}$, here $\textrm{Vec}(A_r^i)$ represents flattening the matrix into a vector. Then the fine-grained user intentions on the relation $r$ for $i$-th head is calculated as:
	\begin{equation}
	\begin{aligned}
	g_i(r, u) &= S^{l-1}_i(e')^{\textbf{T}} A_r^i T^{l-1}_i(e)\cdot\frac{\gamma_{<r, u>}^i}{\sqrt{d}}.
	\end{aligned}
	\end{equation}
	Finally, the overall user fine-grained intention on entity relation $r$
	is calculated by concatenating $k$ heads together:
	\begin{equation}\label{multi_head_relation_attn}
	G(r,u) = \mathop{\rm Softmax}_{N(e)}({\rm Concat}(g_1,...,g_k))[e'].
	\end{equation}
	Here the ``Softmax" is the normalization of the source nodes, and ``[$e'$]" denotes the $e'$-th element. By this, user fine-grained intentions on relations $G(r,u)$ are learned with the guidance of user inherent preferences $\bm{U}_{id}$.

	Similar to the attention calculation procedure, we model the information of source nodes message using $k$ heads projections and concatenated the multi-head information like Eq~\ref{multi_head_relation_attn}:
	\begin{equation}
	f_i(e') = M_r^i W_i^M h^{l-1}(e'),\ \ \ \ \  F(e') = {\rm Concat}(f_1,...,f_k), 
	\end{equation}
	where $M_r^i\in\mathbb{R}^{\frac{d}{k}\times\frac{d}{k}}$ and $W_i^M\in\mathbb{R}^{\frac{d}{k}\times d}$ denote the $i$-th head message matrices.
	
	\textls[-16]{Then we aggregate the message from the source nodes to the target node with the guidance of user inherent attention $G(r, u)$, which finally leads to personalized entity representations and captures the customized and fine-grained user intentions:}
	
	\begin{equation}
	h^l_*(e) = \sum_{\forall e'\in N(e)}(G(r, u)\cdot F(e')).
	\end{equation}
	
	Finally, 
	the entity embedding is finally updated by residual connection~\cite{he2016deep}:
	\begin{equation}
	h^l(e) = \sigma(W^A h^l_*(e)) + h^{l-1}(e),
	\end{equation}
	where $W^A\in\mathbb{R}^{d\times d}$ is the aggregation matrix  and $\sigma(\cdot)$ denotes the activation functions (in practice, we use GELU~\cite{hendrycks2016gaussian}). In the following, we utilize the last layer's (layer $L$) representation $h^L(e)$ as the entity representation and denote $h^L(e)$ as $h(e)$ for simplification.
	
	
	Now given the user mentioned entities \{$e_1, e_2,...,e_n$\} through the conversations, \textls[-2]{we encode them into fine-grained entity representations} $\textrm{H}_u = (h(e_1), h(e_2),...\\, h(e_n))$,
	where $h(e_j)\in\mathbb{R}^d$ denotes the entity embedding of $e_j$. 
	
	\subsubsection{Turn Encoder}\label{entity-order-ext}
	By the entity encoder, we learn the fine-grained user intentions $\textrm{H}_u$ on mentioned entities. However, these entities are not equally important. In CRS, the importance of entities is also influenced by the temporal factor.
	That is, the entities that appeared in the later turns are prone to be more important than early entities. 
	Motivated by the position embedding technique~\cite{vaswani2017attention,devlin2019bert} in NLP, we take the appearing turns of entities into consideration: 
	\begin{equation}\label{self_attn}
	\bm{\mu}_u^o = \textrm{Attn}(\textrm{O}_u) = {\rm Softmax}(\bm{w}^O_2{\rm Tanh}(W^O_1\textrm{O}_u)),\\
	\end{equation}
	where $o_i = \textrm{Emb}_t(t_i)$ is the turn embedding and $\textrm{O}_u=(o_1,...,o_n)$ is the combination of $o_i$. Here $t_i$ is a scalar, which denotes the appearing turn of the entity $e_i$, and $\textrm{Emb}_t$ is the turn embedding layer. 
	We then use the turn importance~(i.e. $\bm{\mu}_u^o$) of entities to better learn fine-grained user intentions. 
	

	\subsubsection{Fine-grained Intention Encoder}
	Actually, the importance of entities is also influenced by the entities themselves. Thus we calculate the self-importance of entities like Eq.~\ref{self_attn}: $\bm{\mu}_u^r = \textrm{Attn}(\textrm{H}_u)$. Finally, we calculate the user representations $\bm{p}_u$ in terms of entity and turn importance, and recommend items according to user intentions:
	\begin{equation}\label{user_intention}
	\bm{p}_u = \frac{1}{2}(\bm{\mu}_u^r + \bm{\mu}_u^o)\textrm{H}_u, \ \ \ \ \ \ 
	\bm{p}_{rec}(i) = {\rm Softmax}(\bm{p}_u\widetilde{\textrm{H}})[i],
	\end{equation}
	\textls[0]{where $\widetilde{\textrm{H}}$ is the embedding matrix of the whole items and $i$ is the index of items.}

	\subsection{Customized Dialogue Generation}
	
	\subsubsection{Sequence-to-Sequence Model}
	The seq2seq framework has been verified in NLP and recommendation \cite{bahdanau2014neural,wu2022selective,wu2022personalized}.
	It consists of an encoder that encodes the input utterances into high-level representations and a decoder that generates the responses. Following~\cite{chen2019towards,zhou2020improving}, we leverage the Transformer~\cite{vaswani2017attention} as base architecture.
	
	Given the input utterance $x=(x_1,...,x_{n_c})$ with dialogue history, the encoder extracts information from $x$. Then the decoder receives the encoder outputs and generates a representation $\bm{q}$ at each decoding time step. According to the representation $\bm{q}$, the generator calculates a probability distribution over the whole vocabulary to determine the generated tokens.

	\subsubsection{Multi-Style Generation}\label{multi-style-gener}
	Actually, the speaking style depends on the current user intention. That is to say, when talking about horror films and romance movies, the speaking style varies definitely. Thus we would like to model the speaking styles of users and perform customized generation.
	
	\textls[-10]{First, we pre-define $n_s$ \textit{latent speaking style embeddings} $\textrm{L}=\{\bm{l}_1,...,\bm{l}_{n_s}\}$}$\in\mathbb{R}^{d\times n_s}$. Then for each user, the corresponding styles vary according to current dialogue contexts and user fine-grained intention. Thus the multi-style vocabulary bias $\bm{g}_u$ is learned by proper speaking styles:
	\begin{equation}\label{eq_choosing_style}
	\begin{aligned}
	\bm{\mu}_u^m = \bm{p}_uW^C\textrm{L},\ \ \ \ \ \ \ \ \ \
	\bm{g}_u = \bm{\mu}_u^m\textrm{L}^{\textbf{T}},
	\end{aligned}
	\end{equation}
	where $\bm{p}_u$ is the user fine-grained intentions learned by Eq.~\ref{user_intention} and $W^C$ is the similarity matrix. The selected style embeddings fit the context semantics and user inherent preferences. Finally, we add the vocabulary bias to the original generator to perform personalized utterances generation:
	\begin{equation}
	\bm{p}_{dial}={\rm softmax}(W^G\bm{q}+\mathcal{F}(\bm{g}_u)+\bm{b}),
	\end{equation}
	where $\mathcal{F}: \mathbb{R}^d\to\mathbb{R}^{|V|}$ maps the vocabulary bias vector into a $|V|$-dimension vector, which consists of two fully connected layers.

	\vspace{-1mm}
	\subsection{Customized Model Training}\label{meta-update}
	In previous sections, we describe the details of our model and capture the customized user preferences from the \textit{design of network} perspective. In this section, we will model the customized user preferences from the \textit{training} perspective.
	
	\subsubsection{Training Loss}
	\textls[-16]{A common way to train the parameters is the back-propagation algorithm. 
		For the recommendation part, we could leverage a common cross-entropy loss to train the recommendation part:}
	\begin{equation}
	\mathcal{L}_{rec}= \frac{1}{|\mathcal{U}|}\sum_{u\in\mathcal{U}} \frac{1}{N_u} \sum_{n=1}^{N_u} \log \bm{p}_{rec}(y_{u,n}),
	\end{equation}
	where $y_{un}$ denotes the actual preference of user $u$ and $N_u$ is the number of the movies in which the user $u$ is interested.
	
	For the dialogue generation module, the common training loss is also the cross-entropy loss:
	\begin{equation}
	\mathcal{L}_{dial} = \frac{1}{|\mathcal{U}|}\sum_{u\in\mathcal{U}}\frac{1}{N_u^g}\sum_{n=1}^{N_u^g}\log \bm{p}(s_{n, u}|s_{(n-1), u},..., s_{1, u}, x_{n, u}),
	\end{equation}
	where $N_u^g$ denotes the whole utterance of user $u$, and $s_{n, u}$ is for the gold response tokens.
	
	Nevertheless, it lacks the personality for various users and could not tune the network according to the customized and fine-grained user preferences. Motivated by~\cite{lee2019melu}, we adopt the meta-learning framework~\cite{finn2017model} to learn the communal user preferences and customized user preferences for a specific user. 
	
	\begin{algorithm}[htb]
		\caption{The training algorithm of recommendation and dialogue generation parts.}
		\begin{algorithmic} 
			\REQUIRE ~~\\ 
			The training model: $m=\textrm{rec}$ or $\textrm{dial}$, \\
			$\beta_m$ and $\nu_m$: inner and outer learning rates.
			\STATE Randomly initialized the inner $\theta_m^{\textrm{inner}}$ and outer $\theta_m^{\textrm{outer}}$ params
			\WHILE{not converge}
			\STATE Sample batch of users $\mathcal{U}\sim \mathcal{D}_{tr}$
			\FOR{user $u$ in $\mathcal{U}$}
			\STATE $(\mathcal{D}^{sup}_u, \mathcal{D}^{qu}_u)\sim\mathcal{D}_{tr, u}$
			\STATE $\mathcal{L}_1(u) = \mathcal{L}_{m, {\mathcal{D}_u^{sup}}}(f_{\theta_m^{\rm inner},\theta_m^{\rm outer}})$
			\STATE Inner update: $\phi(u)_m^{\rm inner} \leftarrow \theta_m^{\rm inner} - \beta_m\nabla_{\theta_m^{\rm inner}}\mathcal{L}_1(u)$
			\STATE $\mathcal{L}_2(u) = \mathcal{L}_{m,{\mathcal{D}_u^{qu}}}(f_{\phi(u)_m^{\rm inner},\theta_m^{\rm outer}})$
			\ENDFOR
			\STATE Global update 
			\STATE $\theta_m^{\rm outer} \leftarrow \theta_m^{\rm outer} - \nu_m \sum_{u\in\mathcal{U}} (\nabla_{\theta_m^{\rm outer}}\mathcal{L}_2(u) + \nabla_{\theta_m^{\rm outer}}\mathcal{L}_1(u))$
			\STATE $\theta_m^{\rm inner} \leftarrow \theta_m^{\rm inner} - \nu_m \sum_{u\in\mathcal{U}} (\nabla_{\theta_m^{\rm inner}}\mathcal{L}_2(u) + \nabla_{\theta_m^{\rm inner}}\mathcal{L}_1(u))$
			\ENDWHILE
		\end{algorithmic}
	\end{algorithm}
	
	\subsubsection{Meta Training}
	The meta training includes the inner update and global update. We first define predicting each user's preference as an individual task and sample a set of records as support set $\mathcal{D}_u^{sup}$, while others as the query set $\mathcal{D}_u^{qu}$ for each user. In the inner update, the model updates the inner parameters $\theta^{\rm inner}$~(see Section Parameters Setting for details) to learn customized user tastes, according to the user's unique item-consumptions (i.e., the support set). It takes a single gradient step with inner learning rate $\beta$:
	\begin{equation}
	\phi(u)^{\rm inner} \leftarrow \theta^{\rm inner} - \beta \nabla_{\theta}\mathcal{L}_{\mathcal{D}_u^{sup}}(\mathcal{M}),
	\end{equation}
	where $\mathcal{M}$ denotes the training parameters of the whole model.
	As the inner parameters are updated in different directions for different users, it captures the user customized preferences. The following procedure is global update and its goal is to learn a communal parameters initialization, such that each of the users' customized preferences would be met from the common initialization with a few update steps, i.e., capturing the customized preferences rapidly:
	\begin{equation}
	\theta \leftarrow \theta - \nu \sum_{u\in\widetilde{U}} \nabla_{\theta} \mathcal{L}_{\mathcal{D}_u^{qu}}(\mathcal{M}(\phi(u)^{\rm inner})),
	\end{equation}
	where $\theta$ includes the whole parameters of $\mathcal{M}$, and $\nu$ is the outer learning rate. 
	
	
	\textls[-12]{In practice, we first train the recommender part to learn the entity embeddings and take the customized recommendation. When the recommender part converges, the dialogue generation module is optimized with the guidance of user intentions on entities, and personalized utterances are generated. The detailed training algorithm of recommendation and dialogue generation is shown in Algorithm 1.}
	
	\subsubsection{Parameters Setting}
	\textls[-16]{For the recommendation part, the inner parameters are set to $\theta_{\rm rec}^{\rm outer}=\{{\rm Emb}_u, {\rm Emb}_e\}$, and the outer parameters are $\theta_{\rm rec}^{\rm inner} = \theta_{\rm rec}\setminus\theta_{\rm rec}^{\rm outer}$, where $\theta_{\rm rec} = \{\theta|\theta\in\mathcal{M}_{\rm rec}\}$.
		For the dialogue generation part, The parameters are also divided into two categories: the inner parameters $\theta_{\rm dial}^{\rm inner} = \{\textrm{Encoder}', W^L_j|_{j=1}^{n_s}\}$ and the outer parameters $\theta_{\rm dial}^{\rm outer} = \theta_{\rm dial}\setminus\theta_{\rm dial}^{\rm inner}$, where $\theta_{\rm dial} = \{\theta|\theta\in\mathcal{M}_{\rm dial}\}$. }
	
	\section{Experiment}
	\textls[-16]{In this section, we first introduce the details of our experiments and then answer the following questions: \textbf{RQ1}: How does our CCRS perform on recommendation and dialogue generation compared with the SOTA baselines? \textbf{RQ2}: Does our CCRS capture the fine-grained user intentions? \textbf{RQ3}: Does our CCRS generate human-like responses? \textbf{RQ4}: How do different components of CCRS benefit the performances?}
	
	\subsection{Experimental Setup}
	\textbf{Dataset.} 
	To evaluate the effectiveness of CCRS, we conduct experiments on real-world dataset ReDial.
	It contains 10,006 conversations and 182,150 utterances. The total number of users and movies are 956 and 51699, respectively. 

	\textls[-16]{For meta-learning, we define each task as a user's interactive conversations. Then we group the train, validation, and test sets by the user id, and the ratio of samples is about 8:1:1. For each user, half of the user conversations are used as the query set, and the remaining conversations are used as the support set. During the meta test phase, CCRS is fine-tuned on the support set $\mathcal{D}_{test}^{sup}$ and tested on the query set $\mathcal{D}_{test}^{qu}$. For a fair comparison with baseline methods, when training these models, we add the meta test support set $\mathcal{D}_{test}^{sup}$ into the training data.}
	
	\noindent
	\textbf{Baselines.}
	We consider the following baselines: (1) \textit{Popularity} ranks the items according to the historical recommendation frequencies. (2) \textit{TextCNN}~\cite{kim2014convolutional}: encodes utterances to extract user intent by CNN-based model. (3) \textit{ReDial}~\cite{li2018towards} is a CRS method which adopts an auto-encoder for recommendation. (4) \textit{KBRD}~\cite{chen2019towards} adopts the external knowledge graph DBpedia to enhance the user representations. (5) \textit{KGSF}~\cite{zhou2020improving} incorporates both the semantic KG ConceptNet and the entity KG DBpedia for modeling user preferences. (6) \textit{KECRS}~\cite{zhang2021kecrs} constructs a high-quality KG and it proposes the Bag-of-Entity loss and the infusion loss to better integrate KG with CRS for generation. (7) \textit{RevCore}~\cite{lu2021revcore} collects user reviews on movies to enhance the recommendation and dialogue generation modules.
	
	
	\noindent
	\textbf{Evaluation Metrics.}
	\textls[-18]{For the recommendation task, we evaluate whether it recommends high-quality items. So we adopt Recall@$k$ (following~\cite{zhou2020improving}), MRR@$k$ and NDCG@$k$ (following~\cite{zhou2020towards}) for evaluation ($k = 10, 50$). For the dialogue generation task, we evaluate the performance by the automatic and human evaluations. In the automatic evaluations, we adopt BLEU~\cite{papineni2002bleu} and F1 to estimate the generation quality, and Distinct $n$-gram ($n = 3, 4$) to measure the diversity at sentence level. 
	In the human evaluations, we invite three annotators to score whether the generations are fluent (\textit{Fluency}) and plenty of useful information (\textit{Informativeness}). The range of scores is 0 to 2, and the final result is the average scores of the three annotators.}
	
	\noindent
	\textbf{Implementation Details.}
	\textls[-16]{We implement our approach based on PyTorch framework. For the recommendation part, the entity embedding size is set to 128. We choose the number of entity relation extractor layers $L=1$ and the number of heads $k=4$. The network parameters are initialized by glorot uniform~\cite{glorot2010understanding}. For training, the inner and outer learning rates are set to 0.006 and 0.003, respectively. For the dialogue part, the dimension of word embeddings is set to 300, and the number of styles $n_s$ equals 4. For training, the inner and outer learning rates are set to 0.0003 and 0.001, respectively. For both parts, we use Adam~\cite{kingma2014adam} optimizer with default parameter setting, and gradient clipping restricts the gradients within [0, 0.1].}
	
	\subsection{Overall Performance (RQ1)}
	
	\subsubsection{Recommendation.} The recommendation results on ReDial are shown in Table~\ref{rec_results}.
	From Table~\ref{rec_results}, we have the following observations. 
	(1) First, our CCRS outperforms CRS baselines (ReDial, KBRD, KGSF, KECRS, and RevCore) by a large margin, which shows the superiority of the fine-grained user intentions extraction and meta training framework. With the guidance of inherent user preferences, the fine-grained user intentions are modeled, and MAML further customizes user preferences by the local update. Moreover, though CCRS does not consider ConceptNet and external user reviews, it outperforms KGSF and RevCore significantly, which also shows the effectiveness of our customization recommendation module.
	(2) Second, CCRS beats the non-CRS method TextCNN. The reason is that TextCNN directly learns user representation by the whole history, which suffers from the sparsity and noise of the utterances.
	(3) Finally, KBRD, KGSF, and KECRS perform better than ReDial, which proves the effectiveness of incorporating external knowledge graphs. Besides, KGSF outperforms KBRD, as KGSF leverages an extra word-oriented KG ConceptNet.
	
	\begin{table*}[!th]
		\small
		\centering
		\setlength{\tabcolsep}{3pt}
		\caption{The recommendation results. The marker * indicates that the improvement is statistically significant compared with the best baseline (t-test with p-value $\textless$ 0.05).} \label{rec_results}
		\begin{tabular}{@{}c|cccccc@{}}
			\toprule
			Dataset & \multicolumn{6}{c}{ReDial} \\
			\midrule
			\textbf{Method} & HR@10 & HR@50 & MRR@10 & MRR@50 & NDCG@10 & NDCG@50\\
			\midrule
			Population & 6.47 & 17.68 & 0.0158 & 0.0204 & 0.0346 & 0.0617\\
			Text CNN & 6.57 & 16.51 & 0.0235 & 0.0275 & 0.0425 & 0.0661\\
			ReDial & 11.79 & 30.11 & 0.0551 & 0.0628 & 0.0895 & 0.1273\\
			KBRD & 15.20 & 33.26 & 0.0593 & 0.0677 & 0.0979 & 0.1382\\
			KGSF & 17.00 & 35.72 & 0.0637 & 0.0717 & 0.1072 & 0.1497\\
			KECRS & 11.57 & 31.95 & 0.0573 & 0.0639 & 0.0911 & 0.1291\\
			RevCore & 17.36 & 36.66 & 0.0659 & 0.0741 & 0.1103 & 0.1528\\
			\midrule
			\textbf{CCRS} & $\textbf{19.09}^*$ & $\textbf{38.41}^*$ & $\textbf{0.0717}^*$ & $\textbf{0.0808}^*$  & $\textbf{0.1193}^*$ & $\textbf{0.1618}^*$\\
			\midrule
			w/o TE & 17.26 & 36.58 & 0.0661 & 0.0751 & 0.1092 & 0.1520\\
			w/o RE\&TE & 16.49 & 35.23 & 0.0633 & 0.0720 & 0.1063 & 0.1463\\
			\bottomrule
		\end{tabular}
	\end{table*}
	
	\begin{table*}[!th]
		\small
		\centering
		\setlength{\tabcolsep}{2pt}
		\caption{Evaluation results on dialogue generation. Flu. and Inf. stand for Fluency and Informativeness, respectively. The marker * indicates that the improvement is statistically significant compared with the best baseline (t-test with p-value $\textless$ 0.05).} \label{dial_results}
		\begin{tabular}{@{}c|ccccccc@{}}
			\toprule
			Dataset & \multicolumn{7}{c}{ReDial}\\ 
			\midrule
			\textbf{Method} & BLEU & F1 & Dist-2 & Dist-3 & Dist-4 & Flu. & Inf. \\
			\midrule

			ReDial & 1.213 & 0.183 & 0.089 & 0.393 & 0.798 & 0.83 & 0.96\\
			KBRD & 1.287 & 0.192 & 0.118 & 0.571 & 1.212 & 0.95 & 1.03\\
			KGSF & 1.629 & 0.227 & 0.123 & 0.647 & 1.583 & 1.23 & 1.32\\
			KECRS & 1.088 & 0.125 & 0.078 & 0.351 & 0.761 & 0.85 & 0.99\\
			RevCore & 1.236 & 0.186 & 0.105 & 0.553 & 1.321 & 1.21 & 1.33\\
			\midrule
			\textbf{CCRS} & $\textbf{2.386}^*$ & $\textbf{0.267}^*$ & $\textbf{0.146}^*$ & $\textbf{0.776}^*$  & $\textbf{1.924}^*$ & $\textbf{1.36}^*$ & $\textbf{1.43}^*$\\
			\midrule
			Only-Meta & 2.129 & 0.258 & 0.131 & 0.692 & 1.772 & 1.30 & 1.40\\
			\bottomrule
		\end{tabular}
	\end{table*}
	
	
	\subsubsection{Dialogue Generation.} We also evaluate our CCRS on the dialogue generation task and the main results are listed in Table~\ref{dial_results}.
	From Table~\ref{dial_results}, we have the following observations. 
	(1) First, our CCRS outperforms the baselines significantly. The improvement on BLEU shows that the generation of CCRS is more consistent with the ground truth, and the high value of Distinct n-gram reflects the diversity of CCRS's results.
	(2) Then, from the human evaluation perspective, CCRS also generates the most fluent and informative responses, which are more human-like responses. The main reason is that CCRS considers multi-style generations according to the current content semantics and inherent user preferences. The human-like generations then improve the user experience, and users are more willing to chat with CCRS.
	(3) Besides, compared with ReDial, we can see that the external knowledge graph also contributes to the generation. 
	
	\subsection{\textls[-0]{Qualitative Results on Recommendation (RQ2)}}
	\subsubsection{Fine-grained user intentions on entity relations.}
	\textls[-20]{In this part, we present qualitative examples to show that we capture fine-grained customized user intentions. }
	
	\begin{figure}[htpb]
		\centering  
		\subfigure[Fine-grained user intentions]{
		    \begin{minipage}[t]{0.7\linewidth}
			    \centering \includegraphics[width=0.99\textwidth]{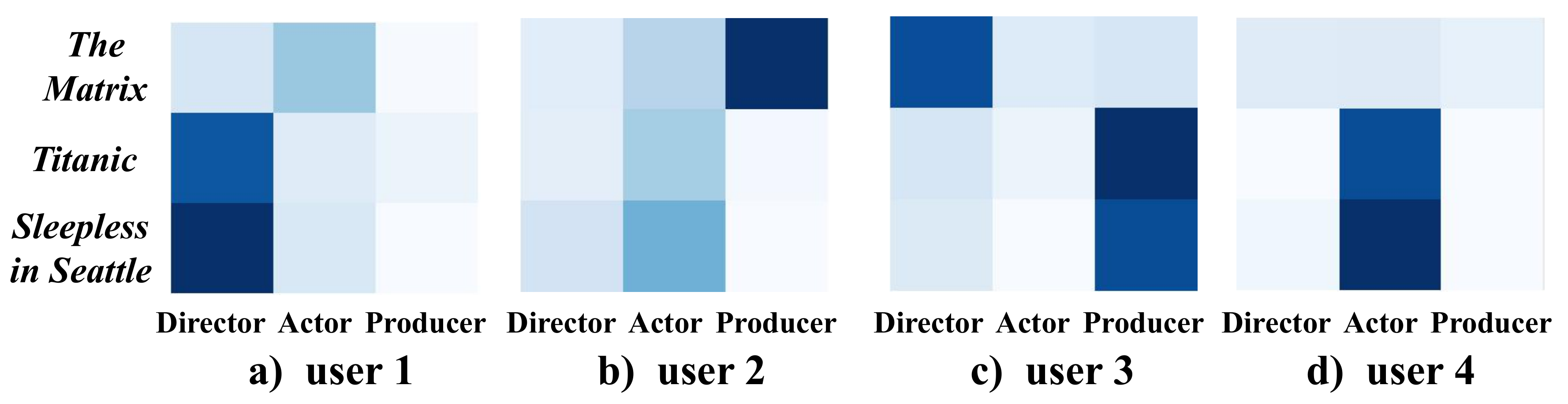}
		        \end{minipage}%
	    }%
	    \subfigure[Turn Importance]{
		    \begin{minipage}[t]{0.28\linewidth}
			    \centering \raisebox{0.9\height}{\includegraphics[width=0.99\textwidth]{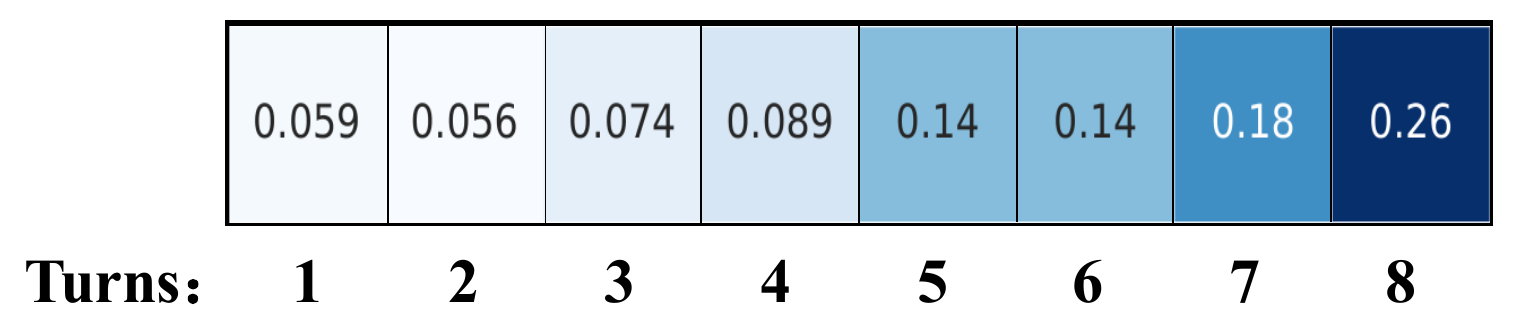}}
		        \end{minipage}%
	    }%
		\caption{The visualization of fine-grained user intentions on different relations (left), and the importance of the dialogue turns (right).}
		\label{rec_visuals}
	\end{figure}

	\textls[-16]{We first choose three movies ``The Matrix", ``Titanic" and ``Sleepless in Seattle" and three different relations ``Director", ``Actor" and ``Producer". Then we randomly pick up four users to 
	show their fine-grained intentions on relations.
	The results are shown in Figure~\ref{rec_visuals} left. (1) Firstly, we can see that the user's preferences on relations vary. 
	For example, user 1 pays more attention to the ``Director" relation, while user 4 cares about ``Actor". 
	So it is necessary to learn the user attention on the entity relation, and we aggregate more information from the ``Director" relation for user 1, while the ``Actor" relation for user 4. 
	(2) Secondly, even for a fixed user, the intentions on different movies also vary. For user 1, he likes the \textit{director} of ``Sleepless in Seattle" and the \textit{actor} of ``The Matrix". 
	(3) Last but not least, in the original corpus, user 4 mentioned that he loves Tom Hanks, which is the star of ``Sleepless in Seattle", and he also likes the movie ``Castaway" which also includes Tom Hanks. 
	These observations show that our CCRS correctly captures the fine-grained user intentions on entity relations and provides high-quality recommendations.}

	\subsubsection{Turn Encoder.}
	\textls[-22]{For the turn encoder, we first conduct an experiment to prove that the appearing turn of the entities contributes to the recommendation performances. Given a user's mentioned entities, we try to rank them purely according to the appearing turn. Then we use the first and last entity's first-order neighbors as the candidate pool. If the candidate pool includes the target movie, this recommendation is viewed as correct. After the calculation, we find the accuracy of the first entity is 2.32\%, while the last entity is 3.50\%. This confirms that the entities near the current turn of the dialogue are more important for learning user preference. 
	Secondly, we visualize the importance of each turn learned by CCRS in Figure~\ref{rec_visuals} right. We observe that with the increase of turns, the attention weights become larger, which helps to recommend high-quality items according to the customized user preferences.}

	\subsection{Qualitative Results on Dialogue (RQ3)}
	\textls[-22]{In this part, we present some qualitative examples to illustrate the personalized generation.
	Firstly, as shown in Table~\ref{dial_case_users}, when talking about different topics of movies, our CCRS could generate multi-style responses according to the current content semantics. While the speaking style of KGSF is monotonous. Thus our CCRS improves the user experience, and the user would feel like chatting with a real person.}
	
	\begin{table}[!t] \small
	\centering
		\caption{The human-like responses for different types of movies.}
		\label{dial_case_users}
		\begin{tabular}{c|ll}
			\toprule
			& User: & I recently watched \textit{The Shining} and it was great. \\
			Topic 1: & KGSF: & Maybe you will also like \textit{Scream}. \\
			Horror film & CCRS: & I'm not sure if you have seen \textit{The Conjuring}. I would say \\
			& & it's pretty horror. It scared me for whatever reason.\\
			\midrule
			& User: & Hello! Can you suggest some movies like \textit{Roman Holiday}?\\
			Topic 2: & KGSF: & \textit{50 First Date} is a good choice.\\
			Romantic Movie & CCRS: & What about \textit{The Notebook}? The characters suffer lots of\\
			& & hardships and get an happy-ending. It tells a moving story!\\
			\bottomrule
		\end{tabular}
		\vspace{-1mm}
	\end{table}
	
	
	\begin{figure}[!htpb]
		\centering  
		\includegraphics[width=0.6\textwidth]{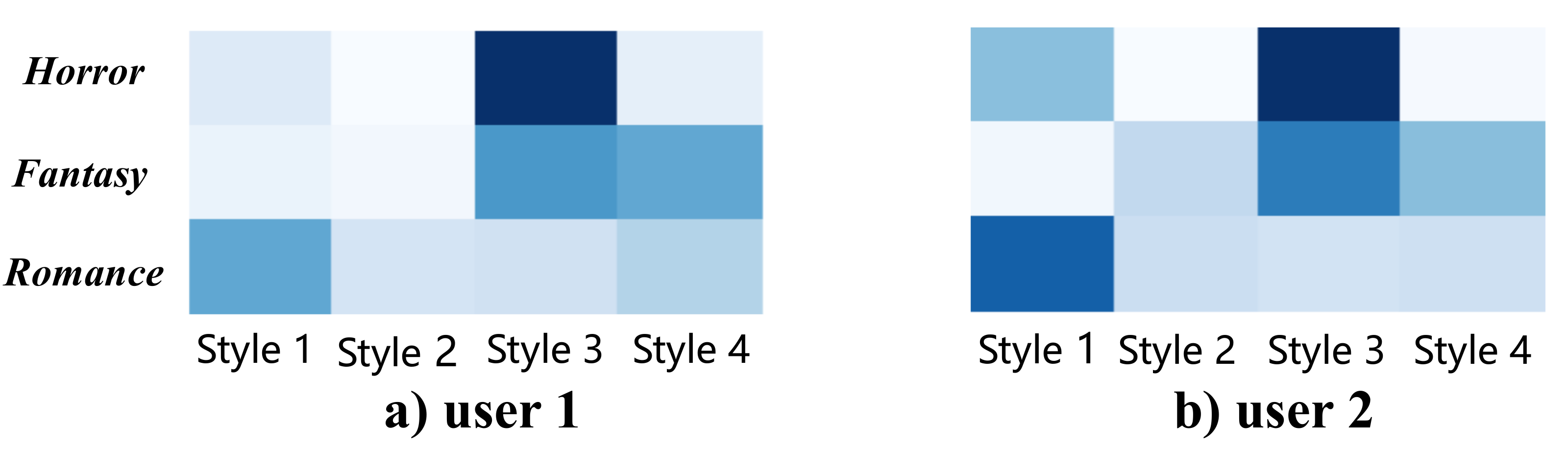}
		\caption{The probability $\bm{\mu}_u^m$ of choosing the different styles. We can see that for different type of movies, the style probability varies.}\label{dial_multi_style}
		\vspace{-0mm}
	\end{figure}
	
	\textls[-16]{Then, we visualize the style distribution vector $\bm{\mu}_u^m$ (see Eq.~\ref{eq_choosing_style}) in Figure~\ref{dial_multi_style}, which controls the choice of the four speaking style embeddings.
	On the one hand, for different types of movies, the style distribution vector varies significantly. 
	For example, for horror film, $\bm{\mu}_u^m$ is prone to choose Style 3, while the romantic movie is Style 1. 
	On the other hand, for the same type of movie, the style distribution is similar among different users, but not identical.
	In a word, CCRS chooses customized style vectors for various kinds of movie topics, which leads to human-like generations.}

	\subsection{Ablation Study (RQ4)}
	\subsubsection{Recommendation.}
	\textls[-16]{In our recommendation part, we propose to learn the fine-grained user intentions, temporal factors of entities, and the meta-learning framework to achieve the customized recommendations. Here we would like to examine the effectiveness of each part and show the results in Table~\ref{rec_results}. 
	For meta training, CCRS w/o RE\&TE denotes training KBRD from the meta-learning perspective (KBRD leverages RGCN to learn the entity representations). CCRS w/o RE\&TE beats KBRD, which shows the effectiveness of meta training. 
	For the remaining two parts, CCRS w/o TE denotes training CCRS with only the entity relation encoder, and it outperforms without entity relation encoder (i.e. CCRS w/o RE\&TE). CCRS incorporates both the entity relation encoder and the turn encoder, and it beats without turn encoder (i.e. CCRS w/o TE) significantly. These results show the usefulness of each part of CCRS. With the help of the fine-grained user intention, the appearing turns of entities, and meta-learning, CCRS achieves good performances.}
	
	\subsubsection{Dialogue Generation.}
	\textls[-12]{In our generation part, we adopt the multi-style generator and the meta training to generate personalized utterances. We also conduct ablation studies to examine the effectiveness of each part. First, the Only-Meta denotes training KBRD with MAML. We can see that MAML significantly improves the quality (BLEU) and diversity (Distinct n-gram) simultaneously. This is not surprising, as we update the network parameters in the user specific direction in the inner update, which leads to the customized generations. Then, CCRS beats Only-Meta, which shows the effectiveness of the multi-style generator. Compared with the Only-Meta, though the improvements on BLEU and F1 scores are marginal, CCRS generates more personalized responses.}
	
	\section{Related Work}

	\textbf{Conversational Recommender Systems} \textls[-0]{Conversational Recommender Systems (CRS)~\cite{zhang2018towards,sun2018conversational,li2018towards,chen2019towards,zhou2020improving,liu2020towards,li2022user} aim to provide high-quality items through the interactive conversations. It mainly consists of a recommender system that identifies the user customized preferences given the item-consumption history and a dialogue system that converses with the users and collects their preferences. Early conversational recommender systems~\cite{christakopoulou2016towards,sun2018conversational} 
	mainly collects the user preference via the pre-defined questions. 
	They pay more attention to recommendation accuracy, and the dialogue system is an auxiliary part, which is implemented by simple or pre-defined patterns.
	Recently, several works focus on building end-to-end CRS models. ~\cite{li2018towards} proposed a standard CRS dataset named ReDial and 
	an end-to-end model which consists of a hierarchical recurrent encoder, a switching decoder, an RNN-based sentiment analysis module, and an autoencoder-based recommendation module.
	Moreover, ~\cite{chen2019towards,zhou2020improving}
	proposed to incorporate the external knowledge to improve the CRS performances. ~\cite{chen2019towards} mainly leverages the knowledge graph (KG) of user mentioned entities, which includes movies, actors, etc., while ~\cite{zhou2020improving}
	incorporates both word-oriented and entity-oriented KGs, and adopts Mutual Information Maximization to align word-level and entity-level representations. Other works focus on selecting proper inter-active action (policy) during the conversations. ~\cite{liu2020towards} constructs a goal sequence which includes question answering, chit-chat, and recommendation phrase. Then they characterized the goal planning process and focus type switch during conversations. ~\cite{zhou2020towards} focused on topic-guided CRS and proposed a topic prediction model. Furthermore, ~\cite{zhou2021crslab} proposed an open-source CRS toolkit CRSLab, which provides a unified and extensible framework for previous CRS works.
	Moreover,  ~\cite{lu2021revcore} proposes a review-enhanced framework, in which user reviews are incorporated to enhance the CRS performances.
	~\cite{liang2021learning} further learns templates automatically for utterances generation.}
	
	\textls[-20]{\textbf{Meta-Learning} Meta-learning is also named learning to learn, which aims to improve new tasks' performance with several similar tasks. 
	The work can be grouped into three clusters, e.g. metric-based~\cite{vinyals2016matching}, model-based~\cite{zhu2022personalized}, and optimization-based~\cite{finn2017model} approaches.
	Recently, meta-learning has also been applied to recommender 
	systems~\cite{pan2019warm,lee2019melu,zhu2021learning,zhu2021warmup,zhu2021transfer,xie2022long} 
	and natural language 
	processing~\cite{dong2020meta}. 
	In this paper, we leverage meta-learning to learn customized conversational recommendation model, and provide personalized service, which is different from previous works.
	To the best of our knowledge, this is the first attempt in adapting meta-learning to conversational recommendation for model customization and improving user experience.}
	
	
	
	\section{Conclusion}
	In this paper, we proposed a novel approach CCRS, which aims to improve the user experience and explore the customization in CRS from three perspectives. For the customized recommendation results, we capture fine-grained intentions by exploring user inherent preference on entity relations and the appearing turn of the entities. For the customized dialogue interactions, we proposed a multi-style generator, which generates responses in customized speaking styles. Finally, for the model training, we adopted the meta-learning framework to enhance the recommendation and dialogue generation via customized model parameters. The extensive experiments show that our CCRS outperforms competitive baselines.
	
	\section*{Acknowledgments}
	This work is also supported by the National Natural Science Foundation of China under Grant (No.61976204, U1811461, U1836206). Zhao Zhang is supported by the China Postdoctoral Science Foundation under Grant No.2021M703273.
	
	%
	%
	%
	\bibliographystyle{splncs04}
	\bibliography{mybibliography}
\end{document}